\documentclass[preprint,endfloats*,superscriptaddress]{revtex4-1}

\usepackage{graphicx}
\usepackage{dcolumn}
\usepackage{bm}
\usepackage{amsmath}
\usepackage{latexsym}

\newcommand{\rfig}[1]{Fig.~\ref{#1}}
\newcommand{\rtab}[1]{Table~\ref{#1}}

\newcommand{\req}[1]{Eq.~(\ref{#1})}

\makeindex

\begin{document}

\title{Magnetoresistance signature of two-dimensional electronic states in Co$_3$Sn$_2$S$_2$}

\author{Jiaji Zhao}
\author{Bingyan Jiang}
\affiliation{State Key Laboratory for Artificial Microstructure and Mesoscopic Physics, Frontiers Science Center for Nano-optoelectronics, Peking University, Beijing 100871, China}
\author{Shen Zhang}
\affiliation{Beijing National Laboratory for Condensed Matter Physics, Institute of Physics, Chinese Academy of Sciences, Beijing 100190, China}
\author{Lujunyu Wang}
\affiliation{State Key Laboratory for Artificial Microstructure and Mesoscopic Physics, Frontiers Science Center for Nano-optoelectronics, Peking University, Beijing 100871, China}
\author{Enke Liu}
\affiliation{Beijing National Laboratory for Condensed Matter Physics, Institute of Physics, Chinese Academy of Sciences, Beijing 100190, China}
\author{Zhilin Li}
\email{lizhilin@iphy.ac.cn}
\affiliation{Beijing National Laboratory for Condensed Matter Physics, Institute of Physics, Chinese Academy of Sciences, Beijing 100190, China}
\author{Xiaosong Wu}
\email{xswu@pku.edu.cn}
\affiliation{State Key Laboratory for Artificial Microstructure and Mesoscopic Physics, Frontiers Science Center for Nano-optoelectronics, Peking University, Beijing 100871, China}
\affiliation{Collaborative Innovation Center of Quantum Matter, Beijing 100871, China}
\affiliation{Shenzhen Institute for Quantum Science and Engineering, Southern University of Science and Technology, Shenzhen 518055, China}
\affiliation{Peking University Yangtze Delta Institute of Optoelectronics, Nantong 226010, Jiangsu, China}

\begin{abstract}
Two-dimensional (2D) Dirac bands and flat bands are characteristics of a kagome lattice. However, experimental studies on their electrical transport are few, because three-dimensional (3D) bulk bands of kagome materials, consisting of stacked 2D kagome layers, dominate the transport. We report a magnetoresistance (MR) study of a kagome material, Co$_3$Sn$_2$S$_2$. Based on analysis of the temperature, magnetic field, and field angle dependence of the resistivity, we obtain a complete anatomy of MR. Besides a magnon MR, a chirality-dependent MR, and a chiral-anomaly-induced MR, the most intriguing feature is an orbital MR that scales only with the out-of-plane field, which strongly indicates its 2D nature. We attribute it to the Dirac band of the kagome lattice.
\end{abstract}



\maketitle

\section{Introduction}

The kagome lattice is expected to host a variety of exotic magnetic and electronic states \cite{Sachdev1992Jun,Lecheminant1997Aug,Balents2010Mar,Han2012Dec,Zhou2017Apr}. Many of them stem from the intrinsic geometric frustration of the lattice \cite{Balents2010Mar,Tang2011Jun,Li2018Nov,Leykam2018Jan,Kang2020Aug}. For instance, a flat band arises because of the localization of electrons in the hexagon due to a destructive interference \cite{Tang2011Jun,Li2018Nov,Leykam2018Jan}. The quench of the kinetic energy in the flat band enhances electronic correlations, giving rise to various unconventional phenomena \cite{Tang2011Jun,Heikkila2011Oct,Neupert2011Jun,Sun2011Jun}. In addition, the lattice symmetry, similar to that of the honeycomb structure of graphene, leads to Dirac bands \cite{Guo2009Sep,Mazin2014Jul}. However, real kagome lattice materials are composed of 2D kagome layers that stack vertically. The interlayer coupling may introduce 3D characteristics and/or disturb the 2D electronic structure \cite{Kang2020Aug}. Consequently, investigation of the kagome physics is often obscured by interlayer coupling effects. It is particularly true in electrical transport when 3D electronic bands are dominant. Therefore, despite spectroscopic evidence of 2D bands \cite{Ye2018,Yin2019,Kang2020Feb,Kang2020Aug,Xu2020a,Liu2020Aug}, it is challenging to investigate their electrical transport properties in kagome lattice materials. Studies of quantum oscillations have revealed an angular dependence of oscillation frequency that suggests 2D Fermi surfaces of a Dirac band, albeit the angle range is too limited to exclude a 3D elliptical Fermi surface \cite{Ye2019Oct,Kang2020Feb,Xu2022Mar}. Huang {\it et al.} have observed an unusual resistivity anisotropy in CoSn and attributed it to a strong in-plane localization of electrons in the flat band \cite{Huang2022Feb}. Still, it was also found that the flat band is dispersive along the out-of-plane direction, indicating a relatively strong interlayer coupling.

Co$_3$Sn$_2$S$_2$ is a ferromagnetic semimetal, consisting of stacked ...-Sn-[S-(Co$_3$-Sn)-S]-... layers, sketched in \rfig{fig.basic}(a). In each layer, Co atoms arrange in a kagome lattice. Near the Fermi level, there are 3D Weyl nodes and nodal lines \cite{Morali2019,Liu2019}, which dominate the electrical transport and manifest various effects, such as the anomalous Hall effect \cite{Wang2018,Liu2018}, chiral anomaly \cite{Liu2018}, chirality-dependent Hall effect \cite{Jiang2021Jun}, thermoelectric effects \cite{Jiang2022Jul}, etc. A scanning tunneling microscopy study has shown a pronounced peak in the density of states at the Fermi level, which was attributed to the kagome flat band \cite{Yin2019}. A flat band due to electronic correlations has also been found in optical spectroscopy \cite{Xu2020a}. Whereas both studies show indications of flat bands, it is not clear whether they are actually 2D. There is so far little transport evidence for 2D energy bands of the kagome lattice.

Here, we perform extensive measurements of magnetoresistance in Co$_3$Sn$_2$S$_2$ single crystals. Through the temperature, magnetic field, and angular dependence of MR, a complete anatomy of MR is carried out. At intermediate and high temperatures, the MR is negative and exhibits an approximately linear dependence on the magnetic field. It is attributed to the field suppression of electron-magnon scattering, which is corroborated by the magnetic field angle dependence of MR. A positive MR appears at low temperatures. It is highly anisotropic regarding the field angle and several contributions can be identified. Besides a chirality-dependent MR and a chiral-anomaly-induced MR that were reported before, there is an MR that follows a roughly quadratic field dependence, consistent with the orbital MR due to a Lorentz force. Most importantly, it scales with the out-of-plane magnetic field. The observed uniaxial MR is strong evidence of 2D electronic transport in Co$_3$Sn$_2$S$_2$.

\section{Experiments}

Two types of Co$_3$Sn$_2$S$_2$ single crystals were investigated in this study. Type I are bulk crystals grown by a chemical
vapor transport (CVT) method \cite{Jiang2021Jun}. Type II are bulk crystals prepared by a flux method \cite{Liu2018}.  All samples were cut and polished. Silver paste was applied to make electrical contacts. Samples were mounted on a motorized rotation stage with a resolution better than $0.02^\circ$. Electrical transport measurements were carried out using a lock-in technique. Four samples in total were measured, among which three are type I and one is type II. All samples show qualitatively the same MR behavior. Data presented in the main text are from B04 and B07, both of which are type I crystals. For B04, the electric current was applied along the $\mathrm{[2\bar{1}\bar{1}0]}$ direction, while it is along the $\mathrm{[01\bar{1}0]}$ direction for B07, as indicated in the inset of \rfig{fig.basic}(b). We adopt a coordinate system, in which the $x$ axis is along the direction of the electric current, while the $z$ axis is normal to the sample plane.

\section{Results and discussion}

\begin{figure}[htbp]
	\begin{center}
		\includegraphics[width=0.9\columnwidth]{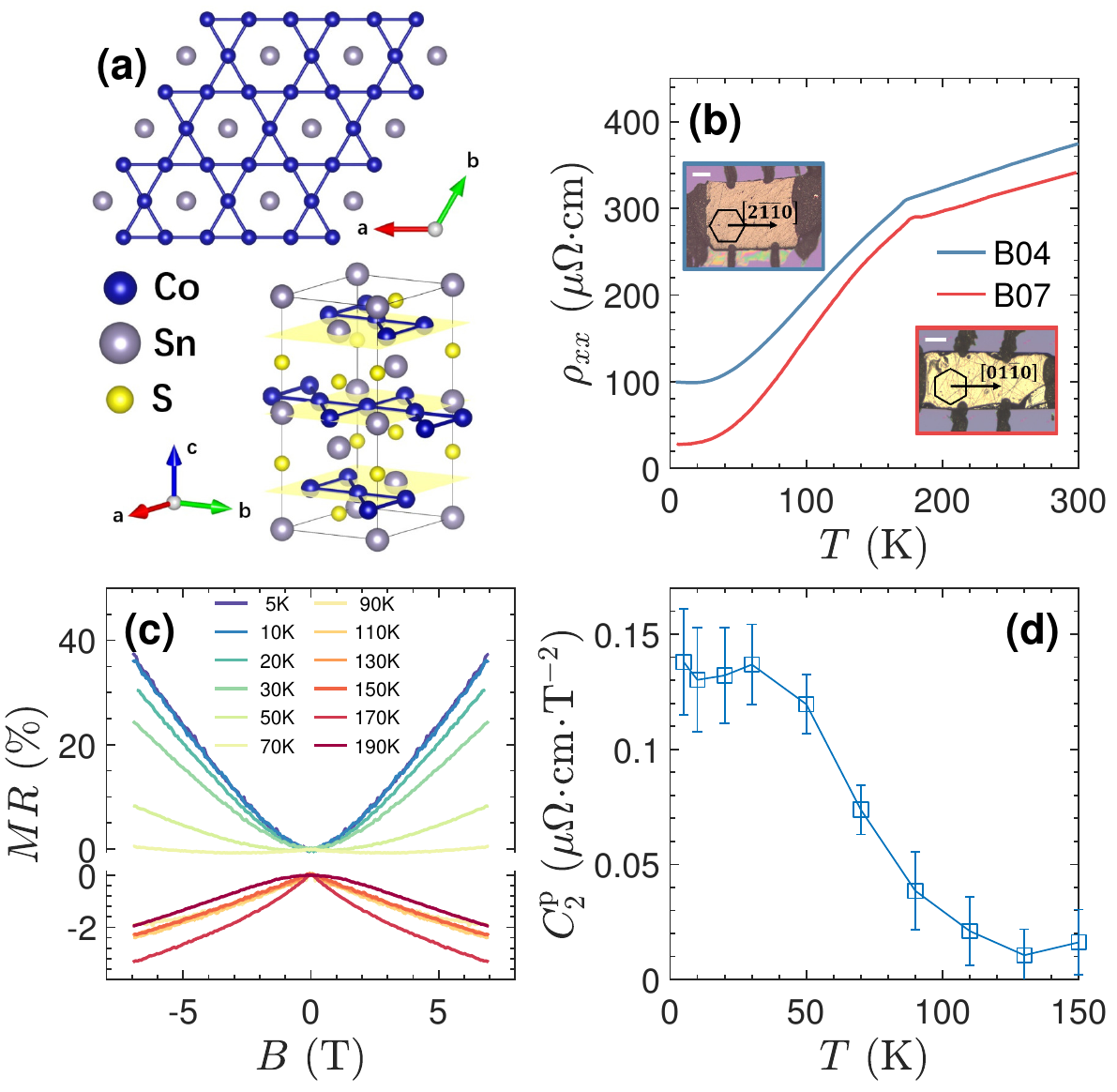}
		\caption{Temperature dependence of the resistivity and MR. (a) Crystal structure of Co$_3$Sn$_2$S$_2$. (b) Temperature-dependent resistivity of two different samples. Their optical images are shown in the inset. The macroscopic shapes and the crystal directions of samples are indicated. Both scale bars are $200\ \mathrm{\mu m}$. (c) MR of B07 under an out-of-plane field at different temperatures. (d) Temperature dependence of the coefficient of the quadratic term in MR.}
		\label{fig.basic}
	\end{center}
\end{figure}

The temperature-dependent resistivity of our Co$_3$Sn$_2$S$_2$ crystals displays a metallic behavior. The residual resistance ratio is relatively large, indicative of high sample quality. A kink in resistivity, characteristic of Co$_3$Sn$_2$S$_2$, appears at the ferromagnetic transition temperature, $T_c=175$ K. MR at different temperatures were measured, shown in \rfig{fig.basic}(c). At low temperatures, the MR is positive and more or less quadratic. With increasing temperature, it diminishes and develops into a negative linear one. The behavior is consistent with previous reports \cite{Yang2020b,Liu2018}. The negative MR persists above $T_c$ and fades away gradually.

Before discussing the positive MR, let us first focus on the negative MR. It is known that weak localization due to electronic phase coherence gives rise to a negative MR \cite{Bergmann1984May}. However, this MR diminishes with temperature, because of the increasing decoherence. This is apparently at odds with the observation that the MR increases with temperature in the intermediate temperature range. In addition, the negative MR of Co$_3$Sn$_2$S$_2$ is roughly linear in magnetic field, seen in \rfig{fig.basic}(c), inconsistent with weak localization. In ferromagnetic metals, a negative MR arises due to the suppression of electron-magnon scattering by magnetic field \cite{Raquet2002,Mihai2008}. The field enhances the magnon gap. The resultant reduction of the magnon population gives rise to the negative MR. Raquet {\it et al.} have considered electron-magnon scattering and calculated the MR in $3d$ ferromagnets \cite{Raquet2002}. The MR was found to approximately depend linearly on magnetic field. Since magnons are thermally excited, the slope of the linearity increases with temperature. As inelastic neutron scattering measurements of Co$_3$Sn$_2$S$_2$ yield a magnon gap of 2.3~meV \cite{Zhang2021Sep}, the magnon population is expected to be strongly suppressed below about $2.3\ \mathrm{meV}/k_\mathrm{B}=27$~K, where $k_\mathrm{B}$ is the Boltzmann constant. Indeed, the magnon MR is not discernible below 30~K. Both the field and temperature dependence of our MR are consistent with magnon scattering.

\begin{figure}[htbp]
	\begin{center}
		\includegraphics[width=0.9\columnwidth]{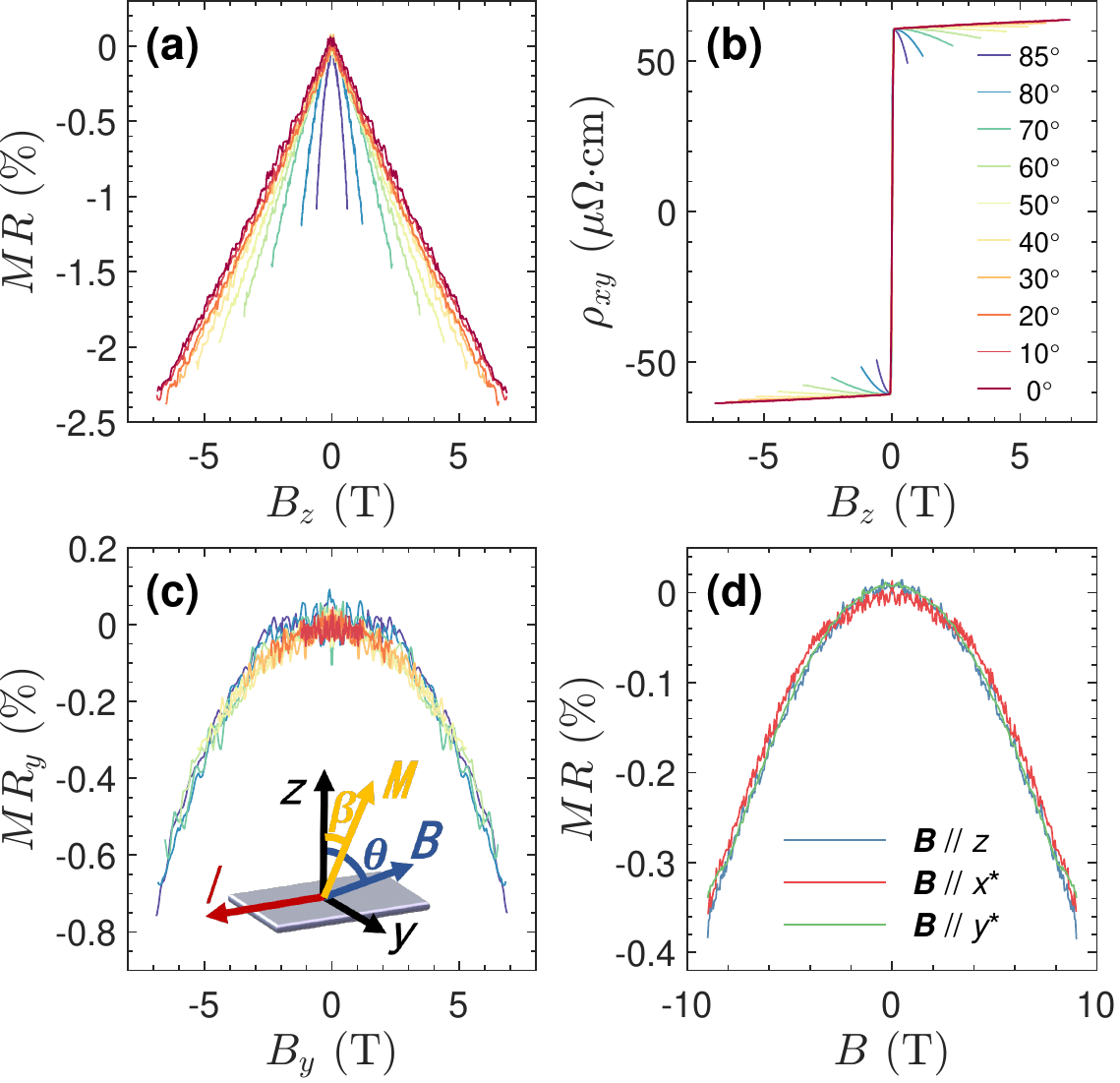}
		\caption{Magnon MR at different field angles. (a) MR of B07 at different angles at 150~K. (b) Hall resistivity at different angles at 150~K. (c) In-plane magnetic field related MR component. The inset shows a schematic of the current and magnetic field direction. (d) MR of B04 at 210~K.\\  *: the field direction is $10^\circ$ out of plane.}
		\label{fig.negativeMR}
	\end{center}
\end{figure}

Since the magnetic field affects the magnon spectrum via the Zeeman energy, it is expected that, in the presence of a strong magnetic anisotropy, the magnon MR is only determined by the field component along the easy axis. We have measured MR at different angles $\theta$ that the magnetic field makes with the $z$ axis in the $y$-$z$ plane. Figure~\ref{fig.negativeMR}(a) shows the MR as a function of the out-of-plane field $B_\perp$ at different angles. Apparently, the MR deviates from the expected simple scaling. The deviation stems from a slight tilt of the magnetization by an in-plane field component. According to the reported measurements, the magnetization of Co$_3$Sn$_2$S$_2$ nearly saturates above the coercive field due to the strong magnetocrystalline anisotropy \cite{Wang2018,Guin2019a,Schnelle2013Oct}. Therefore, the change in magnetization by an in-plane magnetic field is simply a tilt away from the $z$ axis. The tilt angle $\beta$ is a function of $B_\parallel$, i.e., $\sin\beta=f(B_\parallel)$. Since $\beta$ is expected to be small (strong anisotropy), one arrives at $\sin\beta\propto B_\parallel$ under the linear approximation. Such a small tilt is supported by the reduction of the anomalous hall effect, which is proportional to $M_z$, with increasing in-plane field, as seen in \rfig{fig.negativeMR}(b). The direction of the magnetization is sketched in the inset of \rfig{fig.negativeMR}(c). The magnon MR can be expressed as $\Delta\rho(\boldsymbol{B}) \propto B \cdot M \cos(\theta-\beta)=M(B\cos\theta\cos\beta+B\sin\theta\sin\beta)$. When $\beta$ is small, $\cos\beta\approx 1$. Therefore, one has
\begin{equation}
\Delta\rho(\boldsymbol{B}) = C_1 B_z + C_2 B_y^2, 
\label{eq.magnon.MR}
\end{equation}
where $C_1$ and $C_2$ are coefficients independent of the field angle. Our MR data in the intermediate temperature range can be decomposed into two components, seen in \rfig{fig.negativeMR}(c). One component depends linearly on $B_z$ and the other depends quadratically on $B_y$, in excellent agreement with \req{eq.magnon.MR}. The MR above $T_c$ is also negative and follows a quadratic field dependence, which can be well understood based on electron-paramagnon scattering\cite{Fleury1969Apr,Schindler1971Nov} and the linear field dependence of $M$. There is a slight difference in MR between an out-of-plane field and an in-plane one, due to the magnetic anisotropy in the paramagnetic state\cite{Jensen2003May}. The difference vanishes with increasing temperature. As shown in \rfig{fig.negativeMR}(d), the MR becomes isotropic at 210~K.

\begin{figure}[htbp]
	\begin{center}
		\includegraphics[width=0.9\columnwidth]{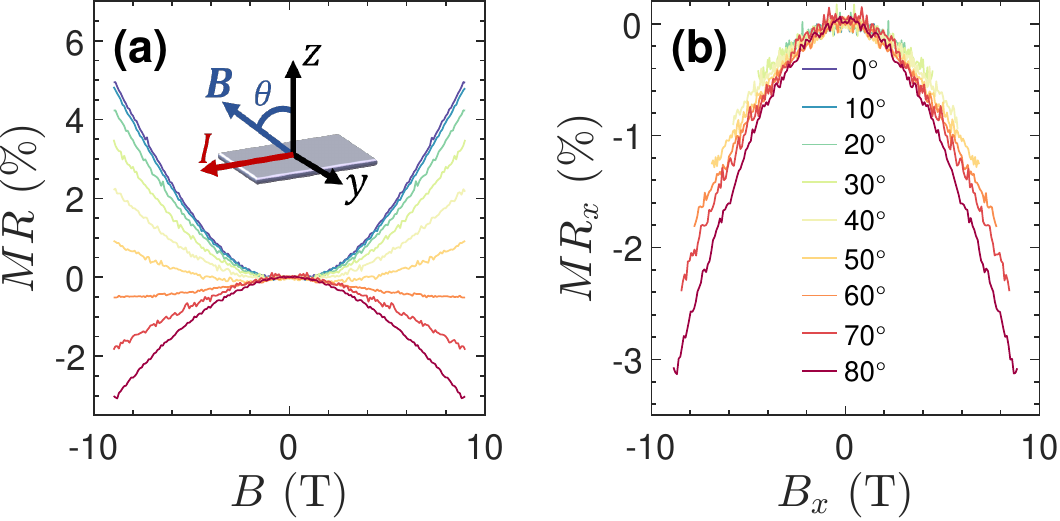}
		\caption{Positive MR for B04 at 20~K. The field is rotated from the $z$ axis towards the current. (a) MR at different angles. (b) In-plane magnetic field related MR.}
		\label{fig.positiveMR.zx}
	\end{center}
\end{figure}

\begin{figure}[htbp]
	\begin{center}
		\includegraphics[width=0.9\columnwidth]{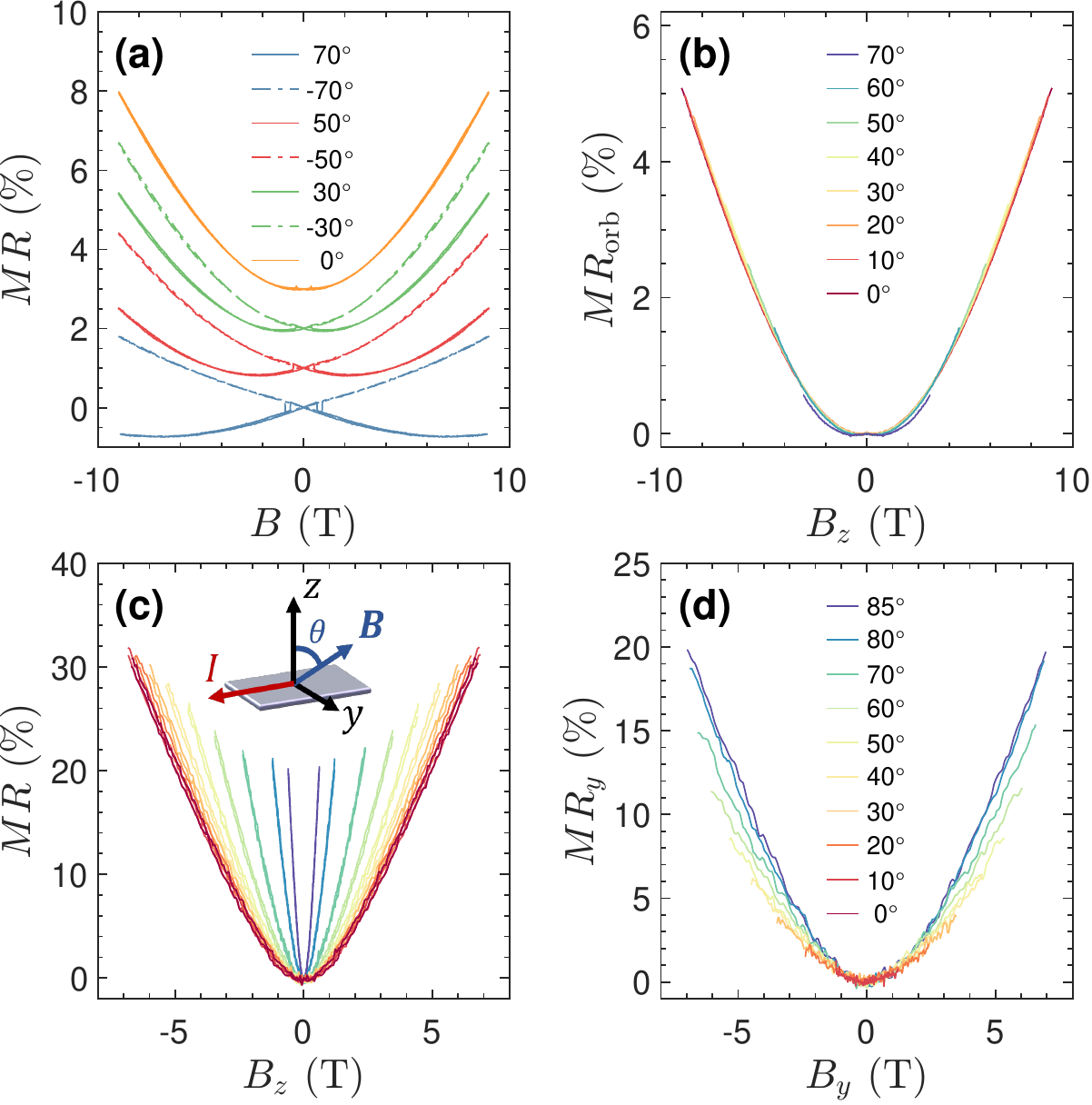}
		\caption{Positive MR at~20 K under a transverse field. The field is rotated from the $z$ axis towards the $y$ axis. (a) MR of B04 at plus and minus angles. (b) MR of B04 as a function of the out-of-plane field after $\theta$-symmetrizing. (c) MR of B07 at different angles. (d) MR of B07 as a function of the in-plane field after removing $MR_\mathrm{orb}$.}
		\label{fig.positiveMR.yz}
	\end{center}
\end{figure}

As temperature decreases, the negative MR is strongly reduced due to the suppression of the magnon population. In the meantime, a positive MR emerges. Since it increases with decreasing temperature, the effect of magnon scattering can be ruled out. Such a crossover from a high-temperature negative MR to a low-temperature positive one is commonly observed in ferromagnetic metals \cite{Raquet2002,Mihai2008}. The low-temperature MR is attributed to the deflection of electron trajectory by the Lorentz force. It usually follows a quadratic magnetic field dependence, which is experimentally observed, as seen in \rfig{fig.positiveMR.yz}. By fitting a parabola to the MR data, we obtain the $B^2$ coefficient $C_2^\mathrm{p}$ and plot it as a function of temperature, shown in \rfig{fig.basic}(d). Above 100 K, the positive MR is marginal and the linear MR becomes dominant. The classical MR due to the Lorentz force is correlated to mobility, according to Kohler's rule \cite{Kohler1938Jan}. With decreasing temperature, the mobility is enhanced owing to diminishing electron-phonon and electron-magnon scattering, as indicated by the temperature-dependent resistivity in \rfig{fig.basic}. The consequent increase of the mean free path leads to the enhancement of MR. Therefore, the temperature dependence of $C_2^\mathrm{p}$ is qualitatively consistent with the classical orbital MR.

The orbital MR should depend on the magnetic field component perpendicular to the electric current and hence vanish when the magnetic field is parallel to the electric current. Figure \ref{fig.positiveMR.zx} shows the MR measured at different angles as the magnetic field is rotated from the $z$ axis towards the direction of the current. The MR diminishes with increasing angle, as expected for an orbital MR. However, it turns negative when $B$ is parallel to the current. A similar negative MR has been observed and attributed to the effect of the chiral anomaly \cite{Liu2018}. The negative MR due to the chiral anomaly is dependent on the parallel field and shows a quadratic field dependence. Therefore, we decompose the measured MR into two parts. The orbital MR is presumably dependent on the out-of-plane field, i.e., $MR_\mathrm{orb}(\theta,B)=MR(\theta=0^\circ,B\cos(\theta))$. The rest, $MR_x(\theta,B)=MR(\theta,B)-MR_\mathrm{orb}(\theta,B)$, turns out to scale approximately with the parallel field. Moreover, $MR_x$ indeed satisfies a quadratic field dependence. Both features of $MR_x$ are consistent with the chiral anomaly.

When the magnetic field rotates in the $y$-$z$ plane, the field is always perpendicular to the current. Therefore, in contrast to the case of $B$ in the $z$-$x$ plane, one would expect that the MR remain unchanged for an isotropic 3D Fermi surface. Figure \ref{fig.positiveMR.yz}(a) shows the MR for B04. Measurements were performed at different field angles $\theta$ with respect to the $z$ axis for the magnetic field in the $y$-$z$ plane. The MR behaves differently at $\pm\theta$. It has previously been shown that the difference is contributed by a chirality-dependent MR originating from the tilting of Weyl cones \cite{Jiang2021Jun}.(Be aware that the chirality-dependent MR and the chiral-anomaly-induced MR are different MR signals.) Since the chirality-dependent MR is an odd function of both the magnetization and the in-plane field, it can be removed by a $\theta$-symmetrizing procedure, $MR_\mathrm{orb}=[MR(\theta)+MR(-\theta)]/2$. Intriguingly, $MR_\mathrm{orb}$ exhibits a strong anisotropy,  i.e., it diminishes when the field is tilted towards the $y$ axis. If the MR traces are plotted as a function of the out-of-plane field $B_\perp$, they all collapse onto a single curve.

To check if the uniaxial MR is universal in Co$_3$Sn$_2$S$_2$, three more crystals were measured. One crystal shows essentially the same $B_z$ scaling behavior of $MR_\mathrm{orb}$. In the other two crystals, it seems that the simple scaling fails, shown in \rfig{fig.positiveMR.yz}(c). Basic information of these samples is shown in \rtab{table:basic}. The procedure that is used to extract $MR_x$ is employed, such that the MR is decomposed into $MR_\mathrm{orb}$ and $MR_y$. It turns out that the extracted $MR_y$ scales reasonably well with $B_y$, which supports the validity of our decomposing method and confirms that $MR_\mathrm{orb}$ in these crystals also scales with $B_z$. Furthermore, the fact that the current in B07 is along the $\mathrm{[01\bar{1}0]}$ (different from the current direction in B04) indicates that the uniaxial MR is independent of the crystalline direction. Two out of four crystals show a nonzero $MR_y$, which follows a power law of $B_y$ with a power index of $\sim 1.5$. $MR_y$ is correlated with neither the direction of the current nor the type of crystals. The difference in the Fermi level or strength of carrier scattering might be responsible for the non-zero $MR_y$ in these two samples. Nevertheless, its origin is currently unclear.

\begin{table}[htbp]
    \centering
    \begin{tabular}{c|c|c|c|c|c}
    \hline
    \hline
        Sample No. & Growth Method & RRR & Residual Resistivity$\mathrm{\ (\mu \Omega\!\cdot\! cm})$ & Current Direction & Size$\ \mathrm{(\mu m^3)}$ \\
    \hline
        B01 & CVT &  3.7 & 117.4 & $\mathrm{[2\bar{1}\bar{1}0]}$ & $1000 \times 770 \times 60$\\
        B04 & CVT & 4   & 99.5 & $\mathrm{[2\bar{1}\bar{1}0]}$ & $1100 \times 1060 \times 60$\\
        B07 & CVT &12  & 27.9 & $\mathrm{[01\bar{1}0]}$ & $1000 \times 450 \times 110$\\
        B10 & flux & 6   & 49.5 & $\mathrm{[2\bar{1}\bar{1}0]}$ & $3500 \times 700 \times 70$\\
    \hline
    \end{tabular}
    \caption{Basic Information of all four samples. RRR stands for residual resistance ratio.}
    \label{table:basic}
\end{table}

The observation that the orbital MR is determined only by the out-of-plane field is surprising. This strongly indicates that the electronic band contributing to $MR_\mathrm{orb}$ is 2D. A kagome lattice is expected to have two characteristic 2D bands, that is, a Dirac band and a flat band. Although an isotropic single band does not have a MR, some extrinsic effects may still give rise to a MR. For instance, inhomogeneities of either carrier density or mobility are known to produce a substantial MR \cite{Parish2003Nov,Jia2014Oct}. This effect is especially strong when the carrier mobility is high, and/or the carrier density is low. Dirac bands usually have high mobility because of their low effective mass. Moreover, many Dirac semimetals in fact exhibit an extremely large MR \cite{Liang2015Mar,Novak2015Jan,Feng2015Aug,Tang2019May}. In contrast, electrons in a flat band are very heavy and of low mobility. As a result, the flat band contributes little to the transport in the presence of other 3D bulk bands. Therefore, the 2D MR is most likely a manifest of the Dirac band.

\section{Conclusion}

In summary, we have studied the magnetoresistance of Co$_3$Sn$_2$S$_2$. Several contributions, including a magnon MR, a chirality-dependence MR, a chiral-anomaly-induced negative MR, and an orbital MR, were identified through the temperature, field, and angular dependence of the resistivity. Most importantly, the orbital MR scales only with the out-of-plane field, which is strong evidence for 2D electronic states. It is likely originated from the Dirac band of the kagome lattice.

\begin{acknowledgements}
This work was supported by National Key Basic Research Program of China (Grants No. 2022YFA1403700, No. 2020YFA0308800, and No. 2022YFA1403800) and NSFC (Projects No. 12074009, No. 11774009, and No. 12204520). Zhilin Li is grateful for the support from the Youth Innovation Promotion Association of the Chinese Academy of Sciences (No. 2021008).
\end{acknowledgements}

\end{document}